\documentclass[aps,prl, superscriptaddress,letterpaper]{revtex4}
\usepackage{graphicx}
\usepackage{amssymb,amsmath}
\usepackage{float}
\usepackage{bm}

\usepackage{setspace}
\usepackage{parskip}
\usepackage{xcolor}
\definecolor{midnightblue}{cmyk}{1,1,0,0.1}
\definecolor{forestgreen}{cmyk}{0.76,0,0.26,0.5}

\usepackage{epsfig}

\usepackage{hyperref}
\hypersetup{
    bookmarks=true,         
    unicode=false,          
    pdftoolbar=true,        
    pdfmenubar=true,        
    pdffitwindow=false,     
    pdfstartview={FitH},    
    pdftitle={My title},    
    pdfauthor={Author},     
    pdfsubject={Subject},   
    pdfcreator={Creator},   
    pdfproducer={Producer}, 
    pdfkeywords={keyword1} {key2} {key3}, 
    pdfnewwindow=true,      
    colorlinks=true,       
    linkcolor=midnightblue,          
    citecolor=magenta,        
    filecolor=midnightblue,      
    urlcolor=midnightblue,          
}

\begin{document}

\title{Interfacial-state coupling induced topological phase transitions in SnTe (110) thin film}

\author{Xiao Li}
\affiliation{Department of Physics, University of Texas at Austin, Austin, TX 78712, USA}

\author{Qian Niu}
\affiliation{Department of Physics, University of Texas at Austin, Austin, TX 78712, USA}
\affiliation{International Center for Quantum Materials, School of Physics, Peking University, Beijing 100871, P. R. China }
\affiliation{Collaborative Innovation Center of Quantum Matter, Beijing, P. R. China }

\begin{abstract}
A defining feature of topological insulating phases is symmetry-protected interfacial Dirac states. SnTe is a representative topological crystalline insulator, of which (110) thin films have two symmetry-unrelated valleys of interfacial states. With the help of valley-contrasting couplings of interfacial states, we design various two-dimensional topological phases in (110) SnTe thin film systems. Our first-principles calculations demonstrate that surface-state coupling strengths of two valleys independently vary with the thickness of the thin film, leading to both two-dimensional topological crystalline insulator and quantum spin Hall insulator. Most interestingly, by constructing a nanoribbon array of SnTe thin film,  edge-state couplings of nanoribbons can further induce topological phase transition between the above topological phases with high tunability, which offers multi-mode quantum transport with potential use in electronic and spintronic devices.
\end{abstract}

\maketitle
\textcolor{forestgreen}{\emph{\textsf{Introduction}.}}---Topological insulating phases,  such as topological insulator (TI) protected by time-reversal symmetry and  topological crystalline insulator (TCI) protected by crystal symmetry, are characterized by symmetry-protected interfacial Dirac states, which have importance theoretical and practical implications on electronics and spintronics \citep{Hasan10,Qi11,Moore10, Ando15}. Materials discovery is crucial and challenging in the study of topological insulating phases. In particular, compared with three-dimensional (3d) counterpart, two-dimensional (2d) TI, also known as quantum spin Hall insulator (QSHI), has only a very few examples that have been experimentally achieved \citep{Konig07, Knez12}. 2d TCI was just recently predicted in (100) thin films of IV-VI semiconductors \citep{Liu14,Ozawa14,Wrasse14, Liu15b}. Therefore, an effective mechanism of designing 2d topological quantum phases is highly desirable.


The coupling between surface Dirac states of 3d topological insulators, e.g. in the Bi$_2$Se$_3$ thin films, modifies the low-energy band dispersion of 2d slab, which may induce the topological phase transition, as proposed in our previous works \cite{Zhou08, Lu10}. 
In this Letter, we develop the mechanism of interfacial-states couplings in a slab with multiple Dirac valleys, in order to design various 2d topological phases in the same system, where the (110) SnTe thin film is taken as an example. Our first-principles calculations demonstrate that 2d TCI and QSHI successively emerge in (110) SnTe thin films as the thickness decreases, which arises from valley-contrasting surface-state couplings. We further construct nanoribbon arrays of topological (110) thin films, where edge-state coupling induced topological phase transitions take place among 2d normal insulator (NI), TI and TCI phases by tuning the widths of nanoribbons or vacuum layers. The emergence of various quantum phases in the same 2d system is especially intriguing. It gives rise to a dissipationless quantized transport of edge channels with multiple stepwise variations, which may lead to novel energy-efficient electronic and spintronic devices. The mechanism of valley-related couplings of interfacial states may also be further used in designing more novel topological phases, such as Chern insulator,  Majorana fermions and parafermions \citep{Fu08,Yu10, Klinovaja14}.

\begin{figure}[h!]
\centering
\includegraphics[width=9cm]{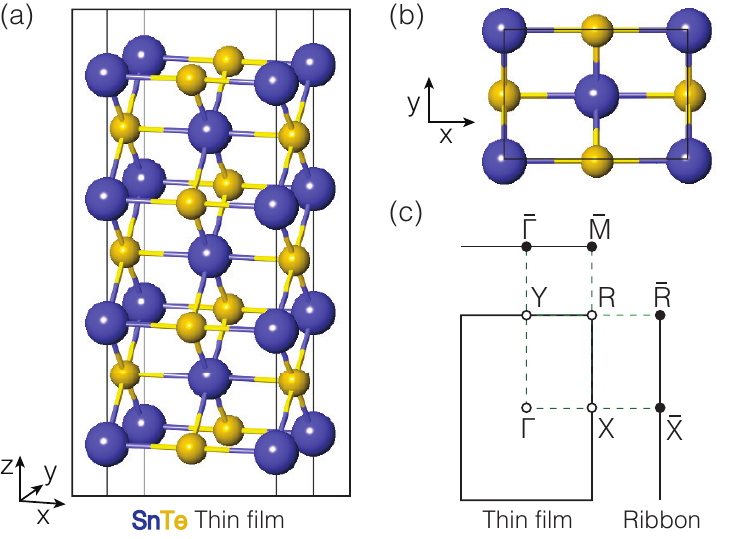}
\caption{Atomic structure and the Brillouin zones of (110) SnTe thin film. (a) Side view and (b) top views of a representative structure of (110) SnTe thin film, which has 7 atomic layers along [110] direction. Blue and yellow spheres stand for Sn and Te atoms, respectively. $x$, $y$ and $z$ axes are along crystallographic [001], [1$\bar{1}$0] and [110] directions, respectively. (c) The Brillouin zones of 2d thin film and its projections along $x$ and $y$ directions. }
\label{fig:1}
\end{figure}

\textcolor{forestgreen}{\emph{\textsf{Surface-state couplings}.}}---Bulk SnTe has a rocksalt crystal structure with (110)-like mirror planes, where two interpenetrating face-centered cubic lattices are formed by tin and tellurium ions, respectively. Under the protection of (110) mirror symmetry, SnTe is a representative topological crystalline insulator. (110) SnTe thin films have two symmetry-unrelated valleys of surface states \citep{Fu12, Liu13}, which may behave differently under the interfacial-states coupling. We therefore investigate geometric and electronic structures of (110) SnTe thin film systems by the first-principles calculations. The detailed calculation method can be found in Supporting Information (S.I.) \cite{SI}.  Fig. \ref{fig:1} shows the relaxed geometric structure and the corresponding Brillouin zone of a (110) SnTe thin film. Each (110) atomic layer of the thin film is a 2d rectangular lattice with a SnTe formula unit in a unit cell. There is an in-plane shift along the diagonal of the rectangular cell by one half of the length of the diagonal and equal vertical intervals between two neighboring layers. Therefore, for thin films with an odd number of atomic layers (e.g. 7 layers in Fig. \ref{fig:1}(a)), the middle layer is still a (110) mirror plane, while the mirror symmetry is absent for thin films with an even number of atomic layers.  We will focus on the films with an odd number of layers, where the presence of mirror symmetry allows of the emergence of 2d TCI and corresponding topological phase transitions.  Besides, the films with an odd number of layers keep inversion symmetry and each ionic site in the mirror plane is an inversion center.

\begin{figure}[h!]
\centering
\includegraphics[width=10cm]{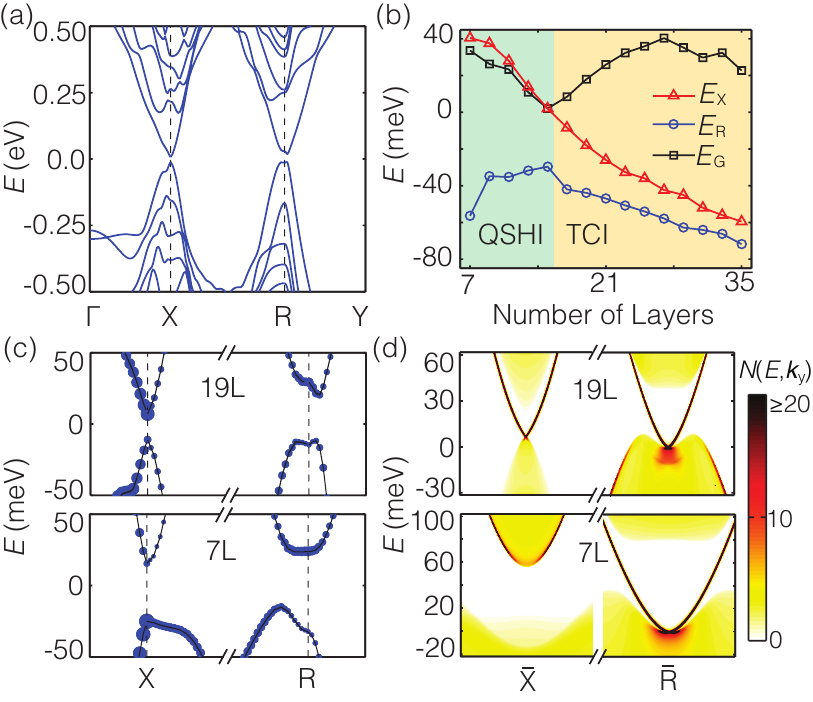}
\caption{Electronic structure of (110) SnTe thin film. (a) Band structure of 19-layer thin film. (b) The evolution of direct band gaps, $E_{\text{X}}$ and $E_{\text{R}}$, at X and R valleys and the global gap, $E_{\text{G}}$, with the thickness of thin film. (c)  Low-energy band structures of 19-layer (upper panel) and 7-layer (Lower) thin films with a fat band representation. The size of blue circle is proportional to the projected weight on Te ions. The bands are shown in the momentum range of $\pm \pi/5$ around X and R valleys. (d) The corresponding energy dispersion at the edge along $y$ axis of the semi-infinite thin film, in the range of $\pm \pi/10$ around $\bar{\text{X}}$ and $\bar{\text{R}}$.}
\label{fig:2}
\end{figure}

Fig. \ref{fig:2}(a) shows a representative band structure of the SnTe thin film. The bands are spin degenerate at each $\bm k$-point of a rectangular Brillouin zone, due to simultaneous time-reversal and inversion symmetries. More importantly, there are indeed two symmetry-distinct valleys around X($\pi$,0) and R($\pi$,$\pi$) points of the $\bm k$-space (Fig. \ref{fig:1}(c)), where the coordinates are given with the lattice constant used as a length unit in each direction.
The low-energy bands of X and R valleys arise from surface states of 3d TCI, given that each valley is a common projection of two band inversion $\bm k$-points of bulk SnTe onto (110) surface \cite{Liu13}.  For a (110) slab of thick enough, two groups of surface states are ideally localized on upper and lower surfaces, respectively. The surface state of each valley on each surface is made up with two interacting coaxial Dirac cones \cite{Wang13}.  The surface state of X valley has mirror-protected gapless Dirac points, which are not located at the high-symmetric X point, but along the mirror projection route, $\Gamma-$X. In contrast,  R valley, away from the mirror projection, is gapped \cite{Liu13,Wang13}. 

As the thickness of the slab decreases below the penetration length of surface states, an additional coupling becomes significant between surface states of the same valley but on different surfaces, which possibly open or tune a band gap associated with the change of the band ordering \citep{Zhou08, Li14, Liu14}. Considering X and R are not related by any symmetry, we expect valley-dependent surface-state couplings and multiple possibilities of band orderings at two valleys that lead to various topological quantum phases in one material.

We study a series of (110) thin films up to 35 atomic layers. 
While the thin film of less than 7 layers is metallic without a global band gap, another film is an insulator, of which the gaps at two valleys and the global gap are shown in Fig. \ref{fig:2}(b), as functions of the (110) atomic layer number, $n$. The gaps of X and R valleys independently vary with thickness, due to changes of surface-state couplings. For X valley, the magnitude of the band gap first decreases to nearly zero ($17\le n\le35$) and then increases ($7\le n\le15$) as the thickness decreases.  In contrast, though there is a similar change trend of the band gap at R valley, the magnitude of the gap is always more than 29 meV. Moreover, the sign of the gaps in Fig. \ref{fig:2}(b) is defined with respect to the band ordering of the atomic limit, which we will explain later.

The closing and reopening of the band gap at X valley is possibly accompanied with a band exchange and a corresponding change of the band topology. To ascertain the changes, we first examine the orbital contributions of band structures of thin films by a fat band representation in Fig. \ref{fig:2}(c). For 19-layer thin film, $p$-orbitals of Te ions mainly contribute to the lowest conduction band states at both valleys, while the highest valence band states have few components from Te ions. The fat band ordering is consistent with the one of bulk SnTe, indicating a band inversion compared with its atomic limit \cite{Fu12}. Given that bulk SnTe is 3d TCI, the inverted bands at both valleys for a thin film suggest a possible 2d TCI. For 7-layer thin film, the orbitals projected on Te ions still contribute to the lowest conduction band states at R valley, while they move to the highest valence band states at X valley. That is, a band exchange between the valence band and conduction band indeed occurs compared with 19-layer thin film, which restores the band ordering of the atomic limit at X valley.  Only one band inversion at R valley may lead to a QSHI in a thin film.  In order to distinguish two band orderings, we define the sign of band gaps at two valleys in Fig. \ref{fig:2}(b), where the negative gap corresponds to inverted bands. 

We further confirm the band topology by the calculations of topological invariants.  The mirror Chern number, $C_m$, used to characterize the mirror-symmetry-protected TCIs, is obtained based on Berry phases  defined by wavefunctions of different mirror eigenvalues in a discretized $\bm k$-space \cite{Fu12, Fukui05, Cai15}. The $\mathcal Z_2$ topological invariant, $\nu$, for identifying TIs is computed by parity eigenvalues of occupied states at time-reversal-invariant (TRI) momenta \cite{Fu07}. For a thin film with $n\ge17$,  we have $|C_m|=2$ and $\nu=0$, demonstrating the system is indeed a 2d TCI. As a result, there are two pairs of counter-propagating gapless Dirac edge states within the band gap at an edge, as shown by edge density of states along $y$ axis in Fig. \ref{fig:2}(d).  Two Dirac points are respectively located at high-symmetric TRI momenta, $\bar{\text{X}}$ ($k_y=0$) and $\bar{\text{R}}$ ($k_y=\pi$), which are respectively projections of the above X and R valleys along $y$ axis (Fig. \ref{fig:1}(c)). In contrast, Dirac points of surface states on (110) and (100) surfaces are not at high-symmetric points \cite{Fu12, Liu13}. These edge states are protected by in-plane mirror symmetry and contribute a robust conductance of 2$e^2/\hbar$.  Similar edge states are also found at an edge along $x$ axis \cite{SI}. On the other hand, for a thin film with $7\le n\le15$, $|C_m|=1$ and $\nu=1$ confirm it as a QSHI. The corresponding edge state has only one Dirac point at $\bar{\text{R}}$ and gives a reduced conductance of $e^2/\hbar$ (Fig. \ref{fig:2}(d)), because only R valley keeps a band inversion. Besides, as shown in Fig. \ref{fig:2}(b), these confirmed 2d TCIs have a maximum global gap of 40 meV when $n=27$, and the maximum global gap of QSHIs is found to be 34 meV with $n=7$. Due to considerable gaps, topological non-trivial quantum transport can be expected at room temperature.

With the help of valley-contrasting couplings of surface states, 2d TCI and QSHI successively emerge in (110) SnTe thin films as the thickness decreases. It is interesting to achieve different 2d topological quantum phases in the same system, which offers a multi-mode quantum transport with a good material compatibility. In contrast, a (001) thin film harbors only 2d TCI phase, with only one independent valley \cite{Liu14}. For a (111) thin film, there is no in-plane mirror symmetry as well as 2d TCI, and the band inversion is not found by first-principles calculations \cite{Shi14, SI}, though it has symmetry-unrelated valleys \cite{Li14,Shi14,Liu15a,Safaei15}.

\begin{figure}[h!]
\centering
\includegraphics[width=9cm]{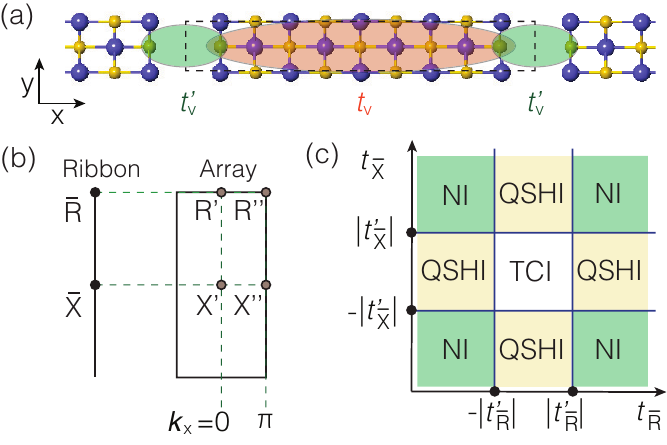}
\caption{Edge-state couplings in SnTe nanoribbon arrays. (a) Top view of a periodic nanoribbon array, with edges of nanoribbons along $y$ axis.  Edge-state couplings through the nanoribbon and the vacuum layer are represented by red and green ellipses, respectively. The unit cell of the array is bounded by dashed lines. (b) When extending from a single nanoribbon to an array, each valley becomes a doublet at TRI points of the 2d Brillouin zone of the array. (c) Topological quantum phase diagram of a nanoribbon array.}
\label{fig:3}
\end{figure}

\medskip
\textcolor{forestgreen}{\emph{\textsf{Edge-state couplings}.}}---
Given the corresponding topological edge states of 2d topological quantum phases, we further design 2d topological phase transitions with higher tunability in (110) SnTe thin film systems, based on edge-state couplings of nanoribbons. We construct periodic nanoribbon arrays of 19-layer (110) thin film, as shown in Fig. \ref{fig:3}(a). Compared with surface-state couplings in a (110) thin film, there are coupling interactions between edge states through the vacuum layer (inter-nanoribbon)  in a nanoribbon array, besides the ones through the SnTe nanoribbon (intra-nanoribbon).  The edge states of the nanoribbons, with gapless Dirac-type dispersions centered at two valleys (Fig. \ref{fig:2}(d)), give rise to the low-energy states at TRI points of 2d Brillouin zone of a nanoribbon array.  Specific to an array composed of nanoribbons with edges along y direction (Fig. \ref{fig:3}(a)),  TRI points, R$'$ and R$''$ (X$'$ and X$''$) are related to ${\bar{\text{R}}(\bar{\text{X}})}$ valleys of the nanoribbon, as shown in Fig. \ref{fig:3}(b). 

 The band gap and the corresponding topology at TRI momenta are completed determined by two kinds of couplings between gapless edge states, which can be well described by a simple low-energy model Hamiltonian of the array, $\mathcal{H}_v=\tau_x(t_v+t'_v \exp(ik_{\perp}))$.
Here, $t_v$ and $t'_v$ are intra- and inter-nanoribbon coupling parameters. $v$ is the valley index to distinguish different Dirac valleys along the momentum parallel to the edge, i.e. $\bar{\text{R}}$ and $\bar{\text{X}}$ valleys in Fig. \ref{fig:3}(b). $\tau=\pm1$ denote two edges of the nanoribbon, respectively.  $k_{\perp}$ is the momentum operator perpendicular to the edge, while the dispersion along the parallel momentum is not taken into account.
We also assume the presence of inversion symmetry for simplicity and then two spins have the same copy of the above term. 
The gaps at $k_{\perp}=0$ and $\pi$,  obtained from $\mathcal{H}_v$, have magnitudes of $2|t_v+t'_v|$ and $2|t_v-t'_v|$, respectively. When $|t_v|= |t'_v|$, the band gap becomes zero at $k_{\perp}=0$ or $\pi$, leading to a possible topological phase transition. 
 By the calculations of parity eigenvalue \cite{Fu07,Li14}, we can obtain the corresponding topological quantum phase diagram of a nanoribbon array in Fig. \ref{fig:3}(c), which also includes both QSHI and TCI. When $|t_v|<|t'_v|$ for both $\bar{\text{R}}$ and $\bar{\text{X}}$ valleys, the array is a 2d TCI; when $|t_v|>|t'_v|$ for both valleys, it is a NI; when $|t_v|<|t'_v|$ for one valley and $|t_v|>|t'_v|$ for the other, it is a QSHI; $|t_v|=|t'_v|$ gives the boundary between different topological phases.

\begin{figure}[h!]
\centering
\includegraphics[width=10cm]{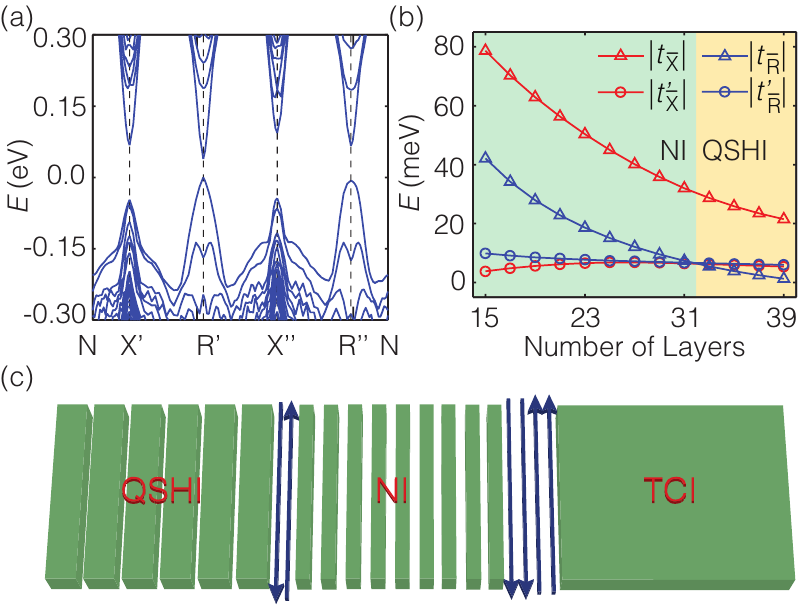}
\caption{Electronic structure of SnTe nanoribbon arrays. A thin film with 19 atomic layers along $z$ direction is adopted. (a) A representative band structure of the array, with nanoribbons of 19 atomic layers along $x$ direction. The coordinates of N is $(\pi/2, \pi/2)$. (b) The evolution of coupling strengths with the width of the nanoribbon. (c) A schematic depiction of  nanoribbon arrays with nanoribbons and vacuum layers of variable width. The nanoribbons and helical edge states are denoted by green blocks and blue arrows, respectively. }
\label{fig:4}
\end{figure}

Under the guidance of topological quantum phase diagram, we then compute the electronic properties of SnTe nanoribbon arrays by tight-binding Hamiltonian based on the Wannier function with first-principles inputs \cite{SI}. A series of periodic arrays of 19-layer (110) thin film is considered, with one nanoribbon of variable width and a fixed vacuum layer of 6.4 \AA\ in a supercell (Fig. \ref{fig:3}(a)).  The model can be seen to be a (110) thin film sheet with a (001) atomic layer removed periodically. The width of the nanoribbon is denoted by the number of (001) atomic layers along $x$ direction, $m$, which is chosen to be odd in order to keep inversion symmetry. Fig. \ref{fig:4}(a) shows a representative band structure of an array, where the spins are degenerate.  Four direct band gaps are located at R$'$, R$''$, X$'$ and X$''$ and determine the band topology. Fig. \ref{fig:4}(b) gives the evolution of the coupling strengths as a function of the width of the nanoribbon. The coupling strengths of both $\bar{\text{R}}$ and $\bar{\text{X}}$ valleys are computed by the band gaps at the above four TRI momenta \cite{SI}. For both valleys, the edge-state couplings through the nanoribbon continuously decrease as the width of the nanoribbon increases, while the couplings through the fixed vacuum layer are nearly unchanged. For the nanoribbon with $m\le31$,  $|t_v|>|t'_v|$ for both valleys, the array is a NI. When the nanoribbon widens more than 31 layers,  $|t_{\bar{\text{R}}}|$ becomes smaller than $|t'_{\bar{\text{R}}}|$, but $|t_{\bar{\text{X}}}|$ is still larger than $|t'_{\bar{\text{X}}}|$ for all array considered, leading to a topological quantum phase transition from NI to QSHI. By further widening the nanoribbon, we expect a switch of coupling strengths at $\bar{\text{X}}$ valley similar to $\bar{\text{R}}$ valley. The array will be a 2d TCI, in consistent with a perfect 19-layer (110) sheet, which, as a 2d TCI, can be regarded as the limit of nanoribbons of infinite width. The band topology of arrays is also confirmed by the fat band and topological invariant. 

By tuning the width of nanoribbons, the topological phase transitions between TCI, QSHI and NI can be achieved. Besides, the width of vacuum layer can be a key parameter to induce topological phase transitions. The edge states decay exponentially when moving away the edge, so do the edge-state couplings between two edges through the vacuum layer. A wider/narrower vacuum layer can therefore move the lines of $|t_v'|$ downwards/upwards in Fig. \ref{fig:4}(b) and have the topological phase transition occur with a wider/narrower nanoribbon. 

It is highly feasible to experimentally realize our proposed topological quantum phases in nanoribbon arrays, with the help of fast-growing lithography techniques \cite{Scappucci09, Walsh09, Son13}.  As shown schematically in Fig. \ref{fig:4}(c), well-patterned nanoribbon arrays by lithography can harbor different topological non-trivial edge states at the phase boundary. The quantum phases and the correponding edge states are determined by the widths of nanoribbons and vacuum layers.
The edge states with conductances of 0, $e^2/\hbar$ and $2e^2/\hbar$ offer a dissipationless quantized transport of multiple stepwise variations, highly tunability and good material compatibility, which has potential use in modern integrated electronics. Moreover, the vaccum layer can be instead by another materials, such as isostructural IV-VI semiconductors, PbTe and GeTe, and some additional interactions from substrates, strain and electric field can also be introduced, which possibly tune the edge-state couplings in a larger range. The phase transition in a nanoribbon array can also apply to another 2d systems with symmetry-unrelated valleys. Even if band gaps at valleys can not be inverted to induce a topological phase transition, tunable valley gaps are desirable in valley-selective optical and electronic applications \cite{Cao12, Li13,Jiang14, Qi15}.

\medskip
\textcolor{forestgreen}{\emph{\textsf{Conclusion}.}} In summary, valley-contrasting couplings between interfacial states are studied in (110) SnTe thin film with symmetry-unrelated valleys, to design 2d topological quantum phases. Both surface-state couplings in a 2d sheet and edge-state couplings in a nanoribbon array lead to TCI and QSHI in the same system by tuning the coupling strengths. In particular, topological phase transitions in an array are expected in a well-controlled way. The multi-mode quantum transport, from time-reversal/mirror symmetry protected helical edge states, gives rise to possible electronic devices with low power consumption and high noise tolerance.


\bigskip

\noindent\textbf{Acknowledgement}
This work is supported by China 973 Program (Projects 2013CB921900), DOE (DE-FG03-02ER45958, Division of Materials Science and Engineering) and Welch Foundation (F-1255).

\end{document}